\documentclass[%
 reprint,
 amsmath,amssymb,
 aps,
]{revtex4-1}

\usepackage{dcolumn}
\usepackage{bm}

\usepackage{amsmath, amssymb}
\usepackage{amsfonts}
\usepackage{braket}

\usepackage[dvips]{graphicx}

\usepackage{ascmac}

\usepackage{fancybox}
\usepackage{type1cm}
\usepackage{comment}

\usepackage{color}

\usepackage{multirow}

\newcommand{\todayd}{\the\year/\the\month/\the\day}
\newcommand{\eq}[1]{\begin{equation} #1 \end{equation}}
\newcommand{\eqa}[2]{\begin{equation} #1 \label{#2} \end{equation}}
\newcommand{\del}{\partial}
\newcommand{\ep}{\epsilon}
\newcommand{\la}{\langle}
\newcommand{\ra}{\rangle}

\newcommand{\bib}{\bibitem}

\newcommand{\mr}{\mathrm}
\newcommand{\lr}{\leftrightarrow}

\newcommand{\balign}[1]{\begin{align} #1 \end{align}}

\newcommand{\lmd}{\lambda}

\newcommand{\lb}{\label}
\newcommand{\nt}{\notag}

\newcommand{\erase}[1]
{
#1
}

\newcommand{\figin}[4]
{
\erase{
\begin{figure}[tb]\centering\includegraphics[width= #1]{#2}\caption{#3}\label{#4}\end{figure}
}
}

\def \({\left(}
\def \){\right)}

\def\rnum#1{\resizebox{0.5em}{\height}{\expandafter{\romannumeral #1}}}
\def\Rnum#1{\resizebox{0.5em}{\height}{\uppercase\expandafter{\romannumeral #1}}}

\begin{document}

\preprint{APS/123-QED}

\title{ Fluctuation Theorem for Partially-masked Nonequilibrium Dynamics}

\author{Naoto Shiraishi}
\author{Takahiro Sagawa}%
\affiliation{%
Department of Basic Science, The University of Tokyo, \\
3-8-1 Komaba, Meguro-ku, Tokyo 153-8902, Japan
}%
\date{\today}

\begin{abstract}

We establish a novel generalization of the fluctuation theorem for partially-masked nonequilibrium dynamics.
We introduce a partial entropy production with a subset of all possible transitions, and show that the partial entropy production satisfies the integral fluctuation theorem.
Our result reveals the fundamental properties of a broad class of autonomous nanomachines as well as non-autonomous ones.
In particular, our result gives a unified fluctuation theorem for both autonomous and non-autonomous Maxwell's demons, where mutual information plays a crucial role.
Furthermore, we derive a novel kind of fluctuation-dissipation theorem that relates nonequilibrium stationary current to two kinds of equilibrium fluctuations.

\begin{description}
\item[PACS numbers]
05.70.Ln, 05.40.-a, 89.70.-a, 87.10.Mn.
\end{description}
\end{abstract}

\pacs{Valid PACS appear here}
\maketitle

\section{Introduction}
In modern nonequilibrium statistical physics, the fluctuation theorem (FT) is significant to characterize the foundation of thermodynamic irreversibility~\cite{Evans,Jarzynski,Crooks,Jarzynski2,Seifert,FTreview}.  
FT has revealed that entropy production is directly related to the probability of the observed trajectory and that of its time-reversal.  
The entropy production is measured by observing the microscopic trajectories, which has been experimentally demonstrated in a variety of systems~\cite{FT-exp,J-exp0,J-exp,Hayashi,Koski}.

In many nonequilibrium systems, however, we are not necessarily interested in all of the microscopic transitions.
A prominent example is Maxwell's demon, which is a composite system of an engine and a memory.
The memory measures the state of the engine and performs feedback control on the engine.
If we calculate the entropy production with the engine alone, the engine apparently violates the FT and the second law of thermodynamics.
Moreover, in many experimental situations with complicated artificial~\cite{inforachet, lock, SEB} and biological~\cite{Hayashi, Toyabe, Ecoli, kinesin, adapt, adapt2} nanomachines, we cannot observe all of the transitions.
If we observe only a part of transitions, we cannot determine the total amount of entropy production.
In such situations, is it still possible to obtain a universal nonequilibrium relation like FT?

In this paper, we reveal the universal property of partially-masked nonequilibrium dynamics.
Let $G$ be the set of all possible transitions between microscopic states, and $\Omega$ be a subset of $G$.
We call transitions in $\Omega$ as {\it observed}, and its complement as {\it masked} (see Fig.~\ref{f:restrict-jump}.(a)).
We then introduce a partial entropy production associated with $\Omega$.
Surprisingly, we can show that the integral FT holds for the partial entropy production, which is regarded as a novel generalization of FT.

The concept of the partial entropy production is straightforwardly applicable to quite a broad class of nanomachines in thermal environment, such as autonomous Maxwell's demons (or bipartite sensing systems)~\cite{Sekimoto, Allahverdyan, Esposito, MJ, Esposito1, Esposito2, HSP, Seifert2, IS, Seifert2014, Jordan2014, BS2}, molecular motors~\cite{Hayashi, Toyabe, kinesin}, ion exchangers~\cite{exchange}, bacterial chemotaxis~\cite{Ecoli, adapt, adapt2}, and single electron boxes~\cite{SEB}.
In order to discuss the power of our result, we show two applications.
First, we apply it to autonomous demons~\cite{Sekimoto,Esposito, Esposito1,Seifert2014,Jordan2014}, which reveals the crucial role of mutual information at the level of stochastic trajectories.
Our approach reproduces the previous results on non-autonomous Maxwell's demons as a special case~\cite{SU2012, SU2012full}.
Moreover, we derive a new kind of the fluctuation-dissipation theorem (FDT) for a pair of transitions.

\figin{7.5cm}{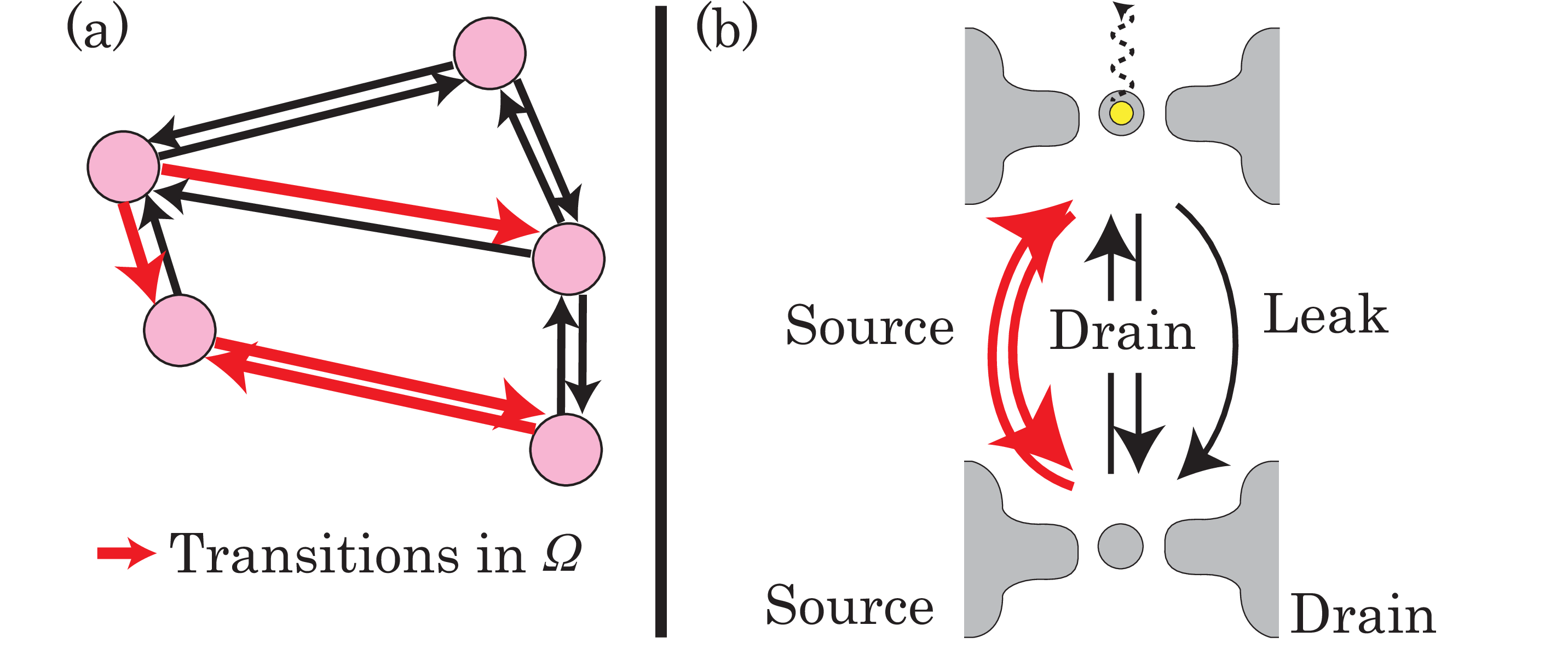}
{(Color online)
(a): Schematic of a Markov jump process, where the circles indicate the microscopic states and the arrows indicate the paths of possible transitions.
The four bold red arrows indicate the observed transitions in $\Omega$ and the eight black arrows indicates its complement.
(b): Schematic of a quantum dot with at most one electron.
Two electron baths, a source and a drain, provide/absorb electrons to/from the dot.
There is also a leak of an electron to outer environment.
We observe only the transfer of electrons between the source and the dot.
Thus, we cannot distinguish the transition associated with the drain from that associated with the leak.
}
{f:restrict-jump}

\section{Total entropy production}

A thermodynamic system obeys continuous-time Markov jump process for time interval $0\leq t\leq T$.
We assume that the number of states of the system is finite.
The transition (i.e., jump) from state $w'$ to state $w$ is written as $w' \to w$, to which we assign transition probability $P(w' \to w ;t)$ that depends on time $t$ in general.
The dynamics of the system is described by the master equation
\begin{equation}
\frac{\partial P(w, t)}{\partial t}=J(w,t):= \sum_{w'} J(w' \to w;t),
\label{master}
\end{equation}
where $P (w, t)$ is the probability of $w$ at time $t$, and $J(w'\to w;t) := P(w',t)P(w' \to w;t) -P(w,t)P(w \to w';t)$ is the probability flux from $w'$ to $w$.  
We assume that the system is attached with a single heat bath at inverse temperature $\beta$.
From the local detailed balance condition, the heat absorbed by the system from the bath during transition $w' \to w$ at time $t$ is given by 
\eq{
Q (w' \to w; t)=-\frac{1}{\beta}\cdot \ln \frac{P(w' \to w;t)}{P(w \to w'; t)}.
}

Let $\Gamma$ be a realized trajectory of the dynamics, in which transitions occur $N$ times at $t= t_1, t_2, \cdots, t_N$.
The state during time interval $t_i \leq t < t_{i+1}$ is denoted by $w_i$ with $t_0:=0$ and $t_{N+1}:=T$.
In particular, the initial and the final states are denoted by $w_0$ and $w_N$, respectively. 
The total entropy production along trajectory $\Gamma$ is then given by 
\eq{
\sigma_{\rm tot}:=-\beta \sum_{i=1}^NQ(w_{i-1}\to w_{i};t_i) +\Delta s,
}
where the stochastic entropy at time $t$ is given by $s(w,t) := - \ln P(w,t)$, and its change is given by $\Delta s:=s(w_N, T)-s(w_0, 0)$.

\figin{3.5cm}{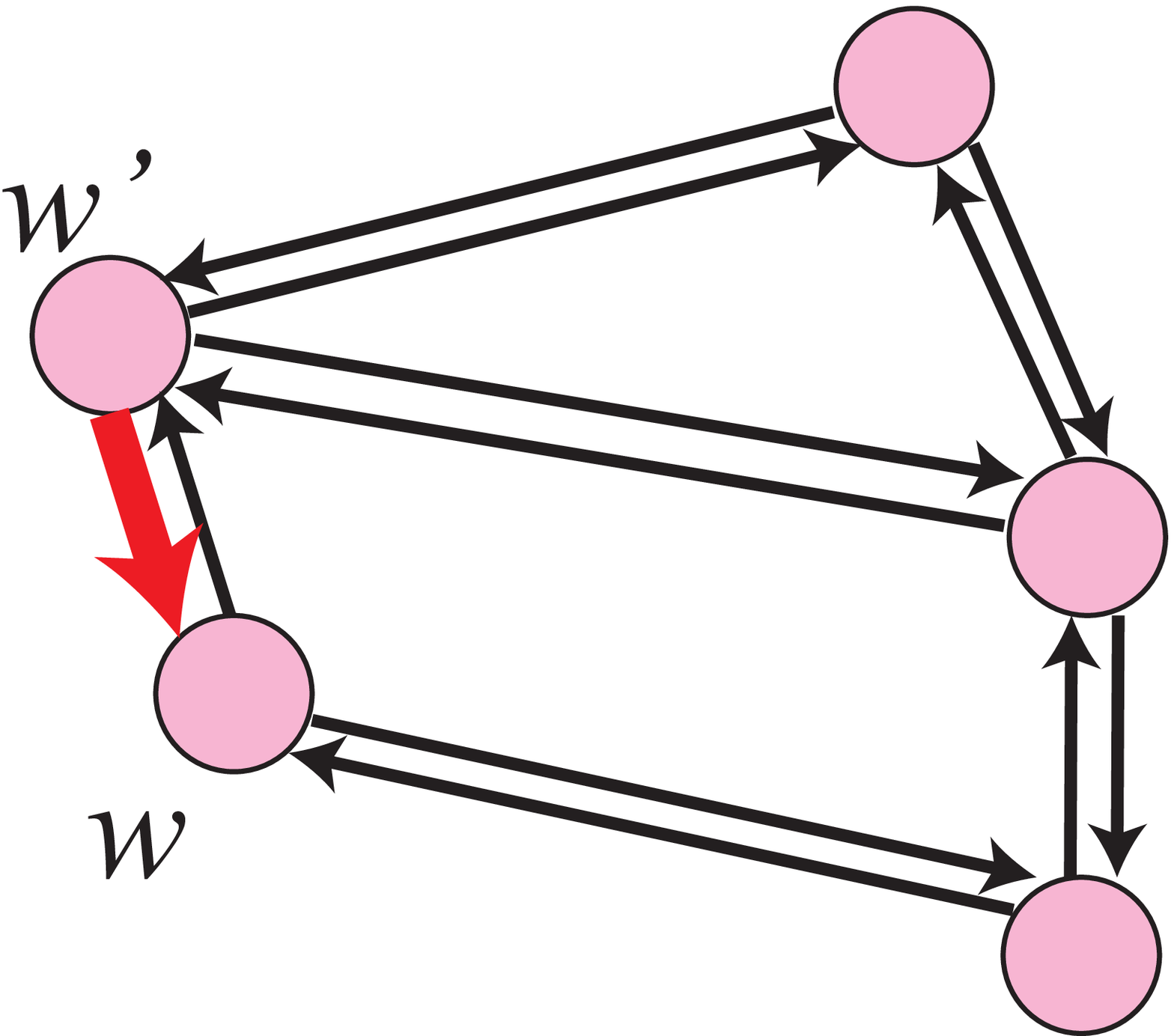}
{(Color online)
An example of the single path $w'\to w$, which is bold and colored by red.}
{f:single}

\section{Main result}

First of all, we define the entropy production associated with a single path of $w'\to w$ (see Fig.~\ref{f:single}):
\eq{
\sigma_{w'\to w} := -\beta Q_{{w'\to w}} + \Delta s_{{w'\to w}}.
\label{observable}
}
The right-hand side (rhs) consists of the following two terms.  
First, $Q_{w'\to w}$ is the heat absorbed by the system during transitions in ${w'\to w}$:
\eq{
Q_{{w'\to w}}:=\sum_{i=1}^{N} Q(w_{i-1} \to w_i;t_i) \delta_{w'\to w}(w_{i-1}\to w_i),
}
where $\delta_{w'\to w}(w_{i-1}\to w_i)$ takes $1$ if $w_{i-1}=w'$ and $w_i=w$, and takes $0$ otherwise. 
Second, $\Delta s_{{w'\to w}}$ is the change in the stochastic entropy induced by the transition ${w'\to w}$:
\eqa{
\Delta s_{{w'\to w}}:=s_{{w'\to w} ,\mr{jump}}-\int_0^T\frac{J(w'\to w;t)\delta _{w(t),w}}{P(w(t),t)}dt,
}{s_w}
where $w(t)$ represents the state at time $t$, and $\delta _{w(t),w}$ takes $1$ if $w(t)=w$ and takes $0$ otherwise.
The first term on the rhs in Eq.~\eqref{s_w} represents the change in the stochastic entropy due to the realized jumps in ${w'\to w}$:
\balign{
&s_{{w'\to w} ,\mr{jump}} \nt \\
:=&\sum_{i=1}^{N}\( s(w_{i},t_{i})-s(w_{i-1},t_{i})\)  \delta_{w'\to w}(w_{i-1}\to w_i).
}
The second term on the rhs in Eq.~\eqref{s_w} represents the change in the stochastic entropy due to the time evolution of the probability distribution induced by transitions in ${w'\to w}$.
The sum of the second term on the rhs in Eq.~\eqref{s_w} for $w'$ equals to the time differential of stochastic entropy:
\eq{
\frac{\partial s(w,t) }{\partial t} = -\sum _{w'}\frac{J(w'\to w;t)}{P(w,t)}.
}

We can then show that the sum of the single-path entropy production for all paths recovers the total entropy production:
\eqa{
\sigma_{\rm tot} = \sum_{w'\to w\in G} \sigma_{w'\to w},
}{sigma-add}
which is a crucial property of the definition~(\ref{observable}). 
By summing up the single-path entropy production over a subset of all paths, we define the partial entropy production with a subset $\Omega \subset G$;
\balign{
\sigma _\Omega &:=\sum _{w'\to w\in \Omega }\sigma_{w'\to w} \nt \\
&=-\beta Q_\Omega +s_{\Omega ,\mr{jump}}-\int_0^T\frac{J_\Omega (w,t)}{P(w,t)}, \lb{sigma-omega}
}
where 
\balign{
Q_{\Omega}&:=\sum_{w'\to w \in \Omega}Q_{w'\to w} \\
s_{\Omega ,\mr{jump}}&:=\sum_{w'\to w \in \Omega}s_{{w'\to w} ,\mr{jump}} \\
J_\Omega (w,t)&:=\sum_{\{w'|(w'\to w)\in \Omega\}} J(w' \to w;t).
}
From Eq.~\eqref{sigma-add}, we can show that 
\eq{
\sigma _\mr{tot}=\sigma _\Omega +\sigma _{\Omega ^\mr{c}},
}
where $\Omega ^\mr{c}$ is a complement of $\Omega$.
In general, if $G$ is divided into $m$ parts $\Omega _1, \cdots ,\Omega _m$, then $\sum_{i} \sigma _{\Omega _i}=\sigma _\mr{tot}$ holds.
Therefore, our formalism enables additive decompositions of the total entropy production; we call this property {\it additivity}.

We here discuss a simple example of the choice of $\Omega$.
Figure~\ref{f:restrict-jump}.(b) shows an experimentally-realizable setup of a quantum dot with two electron baths (the source and the drain)~\cite{SEB}. 
At most one electron is in the dot.
Electrons are provided from these two baths.
In addition, there is a leak of electrons; an electron sometimes escapes from the dot to the outside environment that is regarded as the third bath.
Suppose that we only observe transport of the electrons between the source and the dot.
We set transitions associated with the source to $\Omega$, which is denoted by two bold red arrows in Fig.~\ref{f:restrict-jump}.(b).
Note that we cannot distinguish the transition associated with the drain from that associated with the leak, and thus we cannot calculate the total entropy production from observed data.
Even in such a case, we can calculate the partial entropy production with $\Omega$.

In the above example, there is no backward process of the leak, and therefore, the total entropy production and the partial entropy production with $\Omega ^\mr{c}$ are not well-defined.
However, the partial entropy production with $\Omega$ is still well-defined.
In general, in order to define the partial entropy production with $\Omega$, we only assume that, for any $w\to w'\in \Omega$ with $P(w\to w';t)\neq 0$, the backward transition probability is also nonzero $P(w'\to w;t)\neq 0$.

We stress that it is highly nontrivial whether $\sigma_{\Omega}$ satisfies the integral FT.
However, we indeed have that for any choice of $\Omega$
\eqa{
\langle e^{-\sigma_\Omega} \rangle = 1,
}{restrict}
which is the main result in this paper.

We prove Eq.~(\ref{restrict}) as follows.  
We define another transition rate $P^*$ as
\eq{
P^*(w\to w';t):=
\begin{cases}
P(w\to w';t) &(w'\to w)\in {\Omega} \\
\frac{P(w',t)P(w'\to w;t)}{P(w,t)} & (w'\to w)\notin {\Omega}
\lb{aay-reverse}
\end{cases}
}
with the modified escape rate
\balign{
\lmd ^*(w,t) 
:=\sum_{w'} P^*(w\to w';t) 
=\lmd (w,t)+\frac{J_{{\Omega}^\mr{c}}(w,t)}{P(w,t)}, \lb{lmd}
}
where $\lmd$ is the original escape rate of $P$.
It is easy to show that
\balign{
&P(w\to w';t)e^{\( \beta Q(w \to w' ; t)+s(w,t)-s(w',t)\) \delta_\Omega (w\to w')} \nt \\
=&P^*(w'\to w;t)\frac{P(w',t)}{P(w,t)}, \lb{mid1}
}
where $\delta _\Omega (w\to w')$ takes $1$ if $w\to w' \in \Omega$ and $0$ otherwise.
In addition, $J_\Omega (w,t)+J_{\Omega ^\mr{c}}(w,t)=dP(w,t)/dt$ leads to
\balign{
e^{\int_{t'}^{t''}J_{\Omega}(w,t)/P(w,t)dt}=\frac{P(w,t'')}{P(w,t')}e^{-\int_{t'}^{t''}J_{{\Omega}^\mr{c}}(w,t)/P(w,t)dt}. \lb{mid2}
}
By using Eqs.~\eqref{lmd}, \eqref{mid1}, and \eqref{mid2},  we arrive at our main result
\balign{
&\la e^{-\sigma _\Omega}\ra \nt \\
=&\int d\Gamma P(w_{N},T)\prod_{i=1}^{N}P^*(w_{i}\to w_{i-1};t_{i})\prod_{i=0}^Ne^{-\int_{t_{i}}^{t_{i+1}}\lmd ^*(w_{i},t)dt} \nt \\
=&1.
}

Since Eq.~\eqref{restrict} is valid for any Markov jump systems and any choice of $\Omega$, we obtain many relations in specific situations by applying Eq.~\eqref{restrict}.
In the following, we show two applications.
One is to bipartite systems, which clarifies how information is used in autonomous measurement and feedback.
The other gives a new FDT for a pair of transitions, in which the empirical measure fluctuation plays as an important role as the current fluctuation.

\figin{8.5cm}{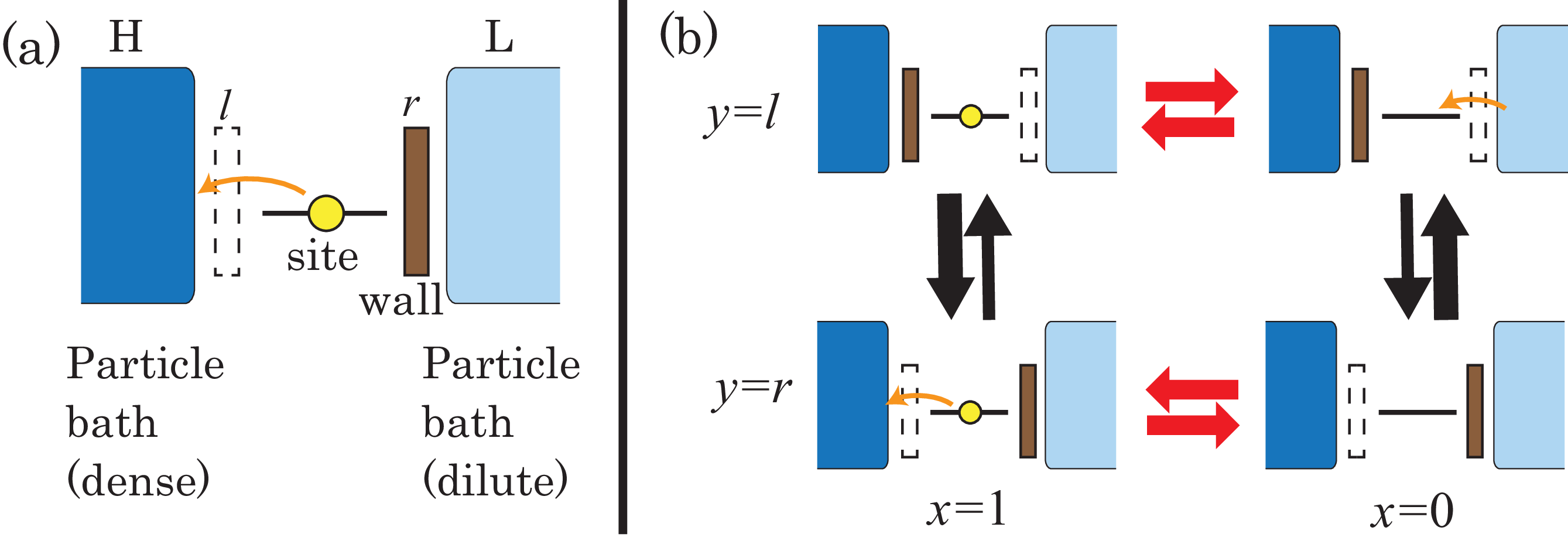}
{(Color online)
(a) Schematic of the autonomous demon, which consists of two baths, a site for single particle, and a wall. 
(b) State space of the model.
If a particle is (is not) in the site, the wall tends to go right (left).
The red arrows indicate transitions in $\Omega$.
With one counterclockwise rotation, one particle is carried from L to H.
}
{f:4state}

\section{Autonomous Maxwell's demons}\lb{s:demon}

We consider a model of autonomous Maxwell's demons, which is a simplification of models discussed in Refs.~\cite{Sekimoto,Esposito, Esposito1}.
We call the system autonomous when the transition rates are time-independent.
Suppose that a particle is transported between two particle baths:  H  with high density and L  with low density  (see Fig.~\ref{f:4state}).
Between the baths, there is a single site where at most a single particle can come in. 
Let $x\in\{0,1\}$ be the number of the particle in the site.
In addition, we consider  a wall that plays the role of the demon.  
The wall is inserted between the site and one of the baths. 
Let $y\in\{l,r\}$ be the position of the wall corresponding to left or right.  
If $y=l$ ($y=r$), the wall prohibits the jump of the particle between the site and the bath H (L).  
The state of the total system is written as $w:=(x,y)$.
Correspondingly, we denote $w_i=:(x_i,y_i)$.
We assume that the probability of $y=l$ is higher (lower) than $r$ if $x=0$ ($x=1$).
Intuitively, the wall measures $x$ and then changes its own state depending on the measurement result, which enables the particles to move from L to H against the chemical potential difference.
However, since the time intervals for measurement processes and that for feedback processes are not separated with each other, the previous results for non-autonomous demons~\cite{SU2012, SU2012full}, in which the mutual information plays a crucial role, cannot apply to autonomous cases.
In more recent works, the role of the mutual information has been clarified for autonomous demons at the level of ensemble average~\cite{Esposito2, Seifert2, Seifert2014, Jordan2014}.
Here, by applying our general result, we will show that the mutual information also plays an important role in autonomous demons at the level of stochastic trajectories, as is the case for non-autonomous demons~\cite{SU2012, SU2012full}.

We introduce the entropy production associated with $x$, $\sigma _x:=-\beta Q_x+s(x_N,T)-s(x_0,0)$, and the mutual information that quantifies the correlation between $x$ and $y$.
The stochastic mutual information between the particle and the wall is given by $I_t(x;y):=\ln \( P(x,y,t)/P(x,t)P(y,t)\)$~\cite{SU2012, SU2012full}, whose ensemble average gives the mutual information~\cite{Cover-Thomas}.
The change in the mutual information associated with the dynamics of the particle is given by
\eq{
\Delta I_x:= I_{x,\mr{jump}}+\int_0^TF_x(x(t),y(t),t)dt.
}
Here, $ I_{x,\mr{jump}}$ represents the change in the mutual information induced by jumps in $x$:
\eq{
I_{x,\mr{jump}}:=\sum_{i=1}^N\( I_{t_i}(x_i;y_i)-I_{t_{i}}(x_{i-1};y_{i-1})\) \delta_{y_i,y_{i-1}} ,
}
where $\delta$ is Kronecker delta.
With notation $J_{x',x}^y(t):=J((x',y)\to (x,y);t)$, $F_x(x,y,t)$ is defined as
\balign{
F_x(x,y,t):=\frac{1}{P(x,y,t)}\sum_{x'}J_{x',x}^y(t)-\frac{1}{P(x,t)}\sum_{y,x'}J_{x',x}^y(t),
}
which represents the change in the mutual information induced by the time evolution of the probability distribution induced by transitions in $x$.
To confirm the meaning of $F_x(x,y,t)$, we transform $F_x(x,y,t)$ into another representation.
By abbreviating $P(x,y,t)$ to $p_{x,y}$, the mutual information such as $I_t(0;r)$ can be regarded as a function with three arguments $p_{0,r}$, $p_{0,l}$, and $p_{1,r}$ such that $I_t(0;r)=\ln (p_{0,r}/(p_{0,r}+p_{0,l})(p_{0,r}+p_{1,r}))$.
The time differential of mutual information is then written as
\eqa{
\frac{dI_t(x;y)}{dt}=\sum_{c\in \{ 0,1\}}\sum_{d\in \{ l,r\}}\frac{\del I_t(x;y)}{\del p_{c,d}}\cdot \frac{dp_{c,d}}{dt}.
}{Idifferential}
It is easy to show that $F_x(x,y,t)$ corresponds to the contribution to \eqref{Idifferential} from the probability flux with $x$:
\balign{
F_x(x,y,t)=\sum_{c\in \{ 0,1\}}\sum_{d\in \{ l,r\}}\frac{\del I_t(x;y)}{\del p_{c,d}}\cdot \sum_{x'}J_{x',c}^d(t).
}
We note that $\Delta I_x$ is also rewritten as
\eq{
\Delta I_x=\int_0^T\iota _x(t)dt,
}
where $\iota _x(t)$ is defined as
\eqa{
\iota _x(t):=\lim_{\Delta t\to 0}\frac{1}{\Delta t}\( I(x(t+\Delta t); y(t))-I(x(t);y(t))\) .
}{iota}
Defining $\iota _y(t)$ in a similar way, we obtain 
\eq{
\iota _x(t)+\iota _y(t)=\frac{dI_t(x;y)}{dt}.
}
Here, the ensemble average of $\iota _x(t)$ is equal to the dynamic information flow given in Refs.~\cite{Allahverdyan, Jordan2014}.

\figin{9cm}{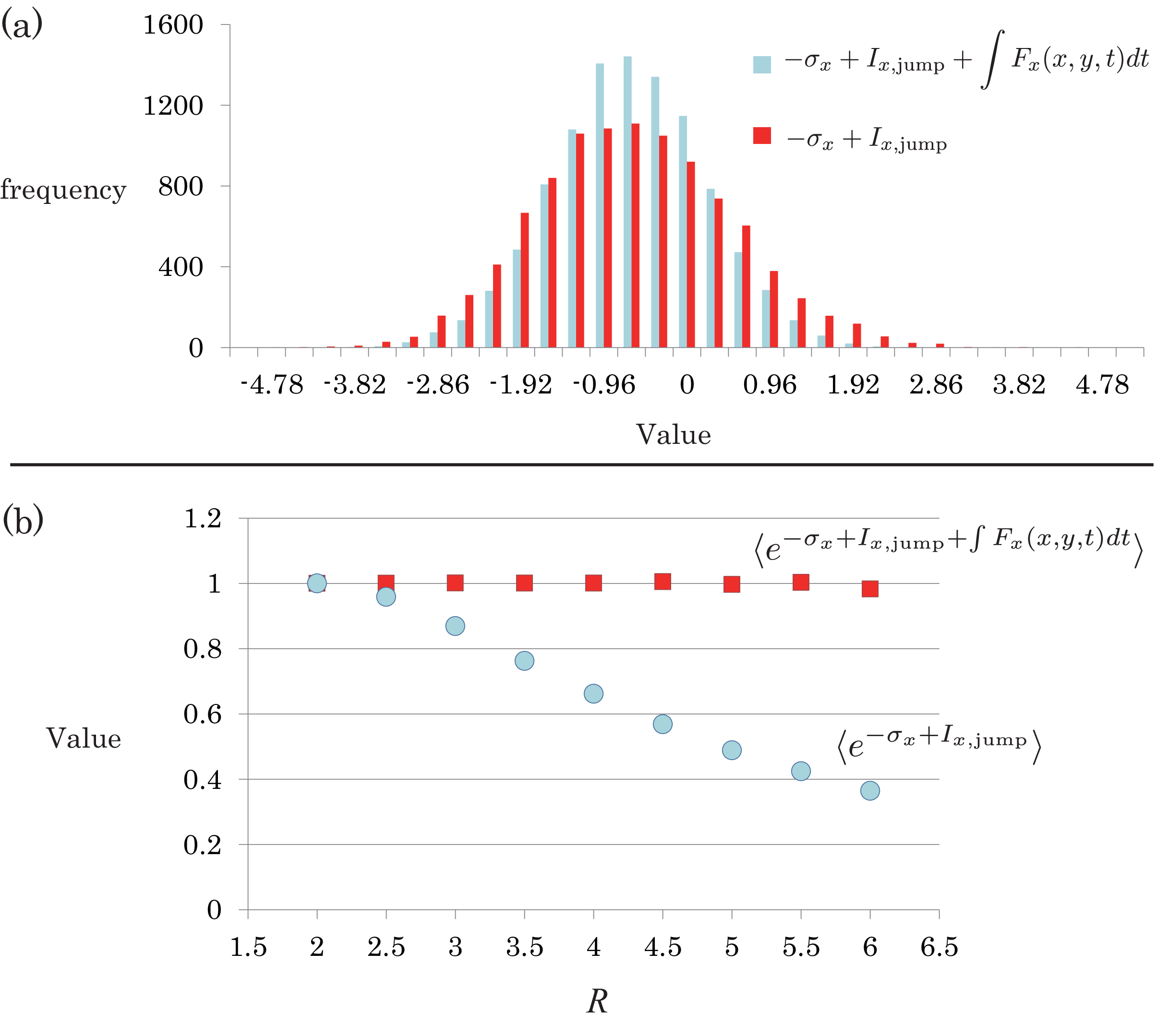}
{(Color online)
Numerical test of Eq.~\eqref{class2-ss}.
(a) A histogram of $-\sigma _x+I_{x, \mr{jump}}$ (blue (light gray) lines) and $-\sigma _x+I_{x, \mr{jump}}+\int F_x(x,y,t)dt$ (red (dark gray) lines) on $R=3.5$ with 10000 trials.
(b) $\la e^{-\sigma _x+I_{x, \mr{jump}}}\ra$ (blue circles) and $\la e^{-\sigma _x+I_{x, \mr{jump}}+\int F_x(x,y,t)dt}\ra$ (red squares) with the change in $R$.
The system is in equilibrium on $R=2$; the larger $R$ is, the larger the stationary flux becomes.
}{f:f-graph}

We now apply Eq.\eqref{restrict} to this model.
We set $\Omega$ to transitions in $x$ (i.e., $\Omega := \{ (0,r) \rightleftharpoons (1,r), (0,l) \rightleftharpoons (1,l) \}$).
Then, $Q_\Omega$ describes the heat absorbed by the particles (i.e., $Q_x=Q_\Omega$).
We also obtain
\balign{
&-\sum _{w'\to w \in \Omega}s_{w'\to w ,\mr{jump}} \nt \\
=&I_{x,\mr{jump}}-\sum_{i=1}^N\( s(x_i,t_i)-s(x_{i-1},t_i)\) \nt \\
=&I_{x,\mr{jump}}-s(x_N, T)+s(x_0,0)-\int_0^T\frac{\sum_{y,x'}J_{x',x(t)}^y(t)}{P(x(t),t)}dt, \lb{sjump}
}
and hence $\sigma _\Omega=\sigma _x-\Delta I_x$.
Then Eq.~\eqref{restrict} reduces to
\eqa{
\la e^{-\sigma _x+\Delta I_x}\ra =1,
}{class2-ss}
in which mutual information contributes to FT on an equal footing with the entropy production associated with the particles.

Notably, for any bipartite system described as $w=(x,y)$ with time-dependent transition rates, Eq.~\eqref{class2-ss} holds with the same derivation.
In this sense, Eq.~\eqref{class2-ss} includes a previously-obtained FT for non-autonomous demons~\cite{SU2012,SU2012full} as its particular case (see Appendix ~\ref{App}).
Thus, Eq.~\eqref{class2-ss} provides a unified view on autonomous and non-autonomous demons, where mutual information is a resource of the entropy decrease of a subsystem.

Using Jensen's inequality, Eq.~\eqref{class2-ss} leads to a second law-like inequality
\eq{
\la \dot{\sigma} _x\ra-\sum_{x,x',y}J_{x,x'}^y(t)\(I_t(x';y)-I_t(x;y)\) \geq 0, 
}
which implies that the entropy production rate of the particles is bounded by the mutual information flow.
This inequality has also been obtained in Refs.~\cite{Allahverdyan, Seifert2014, Jordan2014}.
Note that this inequality does not include any contribution from $F_x(x,y,t)$, because the ensemble average of $F_x(x,y,t)$ is equal to zero.

While the ensemble average of $F_x(x,y,t)$ vanishes, this term is needed in Eq.~\eqref{class2-ss}.
We explicitly show this point with numerical simulation.
Set the parameters $P(1\to 0|\mr{r})=P(0\to 1|\mr{l})=1$, $P(0\to 1|\mr{r})=P(1\to 0|\mr{l})=2$, $P(\mr{r}\to \mr{l}|1)=P(\mr{l}\to \mr{r}|0)=1$, $P(\mr{l}\to \mr{r}|1)=P(\mr{r}\to \mr{l}|0)=:R$, $T=10$, and set the initial state at its stationary state.
We obtain the probability distribution of $-\sigma _x+ I_{x,\mr{jump}}$ (blue (light gray) lines) and that of $-\sigma _x+I_{x,\mr{jump}}+\int F_x(x,y,t)dt$ (red (dark gray) lines) on $R=3.5$.
As shown in Fig.~\ref{f:f-graph}.(a), the variance of the distribution of $-\sigma _x+I_{x,\mr{jump}}+\int F_x(x,y,t)dt$ is larger than that of $-\sigma _x+I_{x,\mr{jump}}$.
Since the tails of the distributions make significant contribution in Eq.~\eqref{class2-ss}, $\la e^{-\sigma _x+I_{x,\mr{jump}}}\ra$ deviates from unity as $R$ increases, whereas $\la e^{-\sigma _x+I_{x, \mr{jump}}+\int F_x(x,y,t)dt}\ra$ stays at unity in agreement with Eq.~\eqref{class2-ss}.

\section{Fluctuation-dissipation theorem}

By expanding our general result \eqref{restrict} around equilibrium, we derive a new FDT for a pair of transitions.
Although the new FDT is general, we here discuss it with a specific example, a simple model of kinesin.
The kinesin conveys an object by consuming the chemical fuel, ATP, with five cyclic steps (see Fig.~\ref{f:kinesin2}).
In this simple model, the kinesin is in equilibrium with stall force.
If the applied force is slightly varied from the stall force, the kinesin is in a linearly nonequilibrium steady state with stationary current.
We will show that the stationary current is characterized by the fluctuations in the equilibrium state.

The perturbation on the transition $w'\to w$ (see Fig.~\ref{f:kinesin2}) is defined as $a:=s(w)-s(w')+\Delta \mu$, where $\Delta \mu$ is the chemical potential difference coupled to reaction $w'\to w$.
If the change in the applied force is of order $\ep$, $a$ is also of order $\ep$.
We then introduce two key quantities.
First, let $N:=\sum_i \delta_{w'\to w}(w_{i-1}\to w_{i})-\delta_{w\to w'}(w_{i-1}\to w_{i})$ be the empirical current from $w'$ to $w$.
The ensemble average of the empirical current equals to the probability flux in the steady state with perturbation $\ep$ such that $\la N\ra =TJ$.
Here, we write $J(w'\to w)$ as $J$ for simplicity.
Next, we define the degree of the fluctuation of the empirical measures for $w$ and $w'$ as $C:=\tau (w')/P(w')-\tau (w)/P(w)$, where $\tau (w):=\int_0^T \delta_{w,w(t)}dt$ is the empirical measure at $w$.
The quantity $C$ indicates how the rate of the empirical measure between $w$ and $w'$, $\tau (w)/\tau (w')$, differs from its ensemble average, $P(w)/P(w')$.
If the empirical measure is equal to the ensemble average, $C$ is equal to 0.
Both the empirical current and the empirical measure are well-studied in the context of large deviation theory~\cite{Derrida, NS, em, mine}.

\figin{6cm}{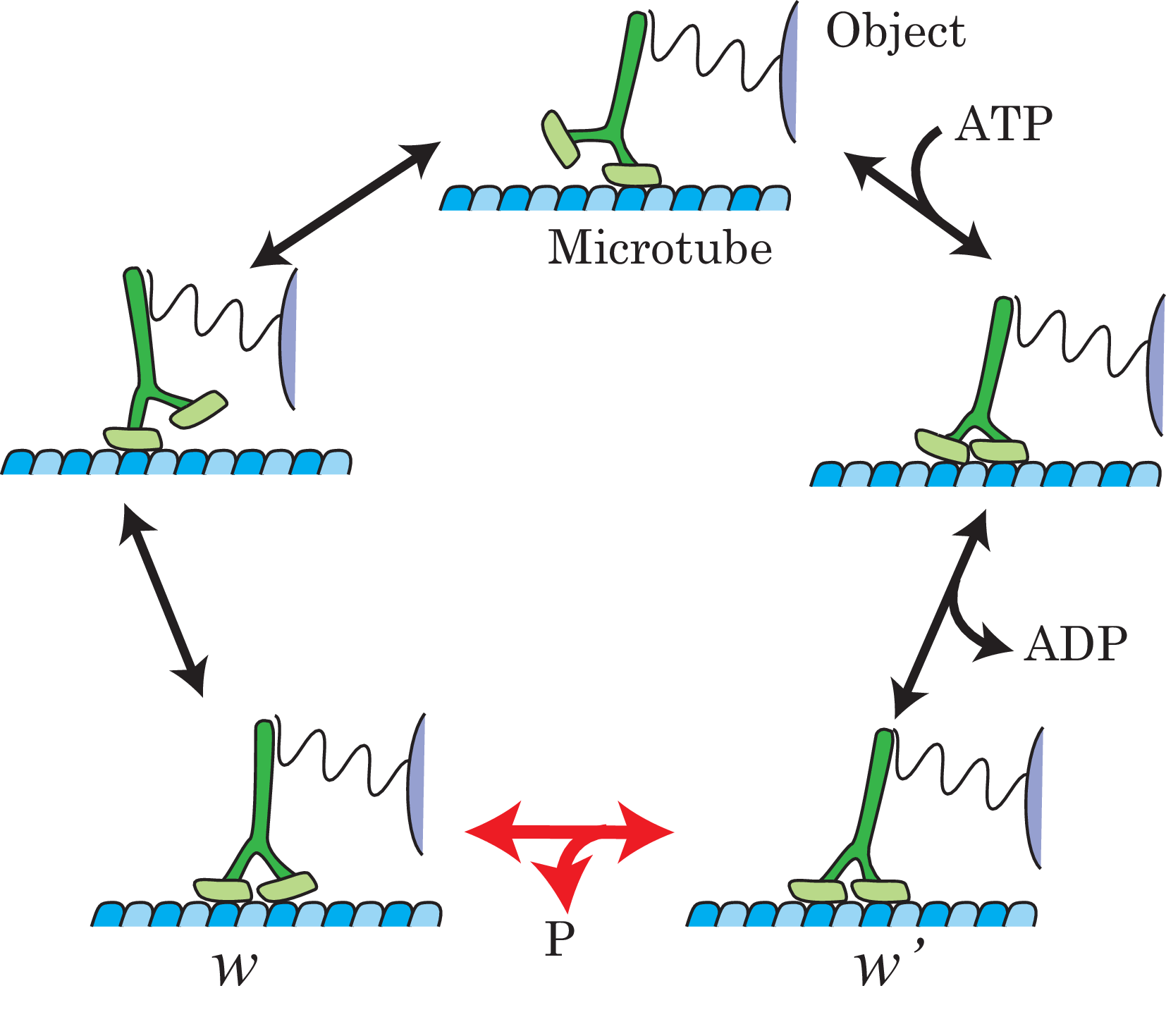}{
(Color online) 
Schematic of the state space of a model of a kinesin.
By consuming ATP, the kinesin conveys an object with five steps.
We focus on the transition $w'\lr w$, which correspond to the bold arrow colored by red, and derive a relation between nonequilibrium current and two equilibrium fluctuations with $w'\to w$.
}{f:kinesin2}

By setting $\Omega$ to the transitions $w'\lr w$, Eq.~\eqref{restrict} is written as $\la e^{-aN+JC}\ra =1$.
Note that $a$ and $J$ are of order $\ep$.
Hence, the above equality is expanded as 
\eqa{
\la -aN+JC\ra +\frac{1}{2}\la (-aN+JC)^2\ra +O(\ep ^3)=0.
}{expand1}
From $\la C\ra =0$ and $\la \cdot \ra =(1+O(\ep ))\la \cdot \ra _0$, Eq.~\eqref{expand1} is transformed into
\eqa{
aTJ=\frac{1}{2}\la (-aN+JC)^2\ra_0 +O(\ep ^3),
}{expand2}
where $\la \cdot \ra _0$ represents the ensemble average in the equilibrium state.
Since $NC$ changes its sign for the time-reversal trajectory, and since in equilibrium a trajectory and its time reversal have the same probability, the cross-term of the rhs, $\la NC\ra_0$, is equal to $0$.
Substituting $\la NC\ra_0=0$ into \eqref{expand2} and taking the equality up to $\ep^2$ order, we obtain a new FDT:
\eqa{
aTJ=\frac{a^2}{2}\la N^2\ra _0+\frac{J^2}{2}\la C^2\ra_0.
}{FDT}
Here, since $\la N\ra _0=\la C\ra _0=0$, $\la N^2\ra _0$ and $\la C^2\ra _0$ represent the current fluctuation and the empirical measure fluctuation in the equilibrium state, respectively.
The obtained FDT~\eqref{FDT} connects the nonequilibrium stationary current $J$ and the two kinds of equilibrium fluctuations.
In contrast to the usual FDT, the empirical measure fluctuation appears in this FDT.
We note that the fluctuation of $C$ is significant in Eq.~\eqref{FDT}, while its ensemble average is zero.

In addition, the condition that $J$ is real number leads to
\eqa{
\frac{1}{T^2}\la N^2\ra _0\la C^2\ra_0\leq 1,
}{CR}
which implies that both of the current fluctuation and the empirical measure fluctuation cannot be large at the same time.

\section{Concluding remarks}

We have derived a novel FT \eqref{restrict} for partially-masked nonequilibrium dynamics.
Applying the general result to specific situations, we can obtain both previous results~\cite{Evans, Crooks, Seifert, FTreview, SU2012, SU2012full} and new relations like Eq.~\eqref{class2-ss}, Eq.~\eqref{FDT}, and Eq.~\eqref{CR}.
Equation~\eqref{class2-ss} clarifies the role of mutual information in both autonomous and  non-autonomous Maxwell's demons.
Equations~\eqref{FDT} and \eqref{CR} show novel features of equilibrium fluctuations with a pair of transition paths.
The single-path entropy production \eqref{observable} is regarded as a building block to construct various thermodynamic relations for Markov processes.
Although we treat only the Markov jump processes in this paper, it is easy to extend our result to the Markov chain and the Langevin dynamics.

We here make a remark on the relationship between our result and a previous work.
Although Eq.~\eqref{restrict} looks similar to an equality obtained by Hartich {\it et.al}. (Appendix A of Ref.~\cite{Seifert2014}), there is a crucial difference between their result and ours.
Their result is a special case of the following equality:
\eqa{
\la e^{\beta Q_{\Omega}-s_{\Omega ,\mr{jump}}-\int_0^T{J_\Omega (w(t),t)}/{P(w(t),t)}dt}\ra =1,
}{Seifert}
where we assumed that, if $w\to w'$ is in $\Omega$ ($\Omega ^\mr{c}$), $w'\to w$ is also in $\Omega$ ($\Omega ^\mr{c}$).
The sign of ${J_\Omega (w(t),t)}/{P(w(t),t)}$ in Eq.~\eqref{Seifert} is opposite to that in Eq.~\eqref{restrict}.
Therefore, the exponent of left-hand side in  Eq.~\eqref{Seifert} does not satisfy the additivity, whereas the additivity is the crucial characterization of our approach.

The partial entropy production given in Esq.~\eqref{observable} and\eqref{sigma-omega} satisfies both of the additivity and the fluctuation theorem.
The additivity implies that the total entropy production can be decomposed into those of the subsets of transitions. 
The fluctuation theorem leads to a variety of new thermodynamic relations.
Therefore, our definition of the partial entropy production is a reasonable way to assign the entropy production to individual transitions.
This approach would enhance our understanding of stochastic thermodynamics at the level of individual transition paths.
For an instance, we have derived a new FDT for a pair of transition paths.
Another possible application of our framework is to biological molecular motors, which are regarded as small heat engines converting the fuel into work~\cite{Toyabe, F1eff, motorbook}.
In this approach, for example, we would be able to reveal a bottleneck process in terms of the thermodynamic efficiency of motors, and its connection to the design principle of the molecular structure.

\acknowledgments

Authors thank Sosuke Ito and Kyogo Kawaguchi for fruitful discussion and useful comments.
Authors also thank Jukka Pekola for helpful suggestions.
This work is supported by JSPS KAKENHI Grant Nos. 25800217 and 22340114.

\appendix

\section{Derivation of the FT for non-autonomous Maxwell's demons from Eq.~\eqref{class2-ss}}\lb{App}

\figin{6cm}{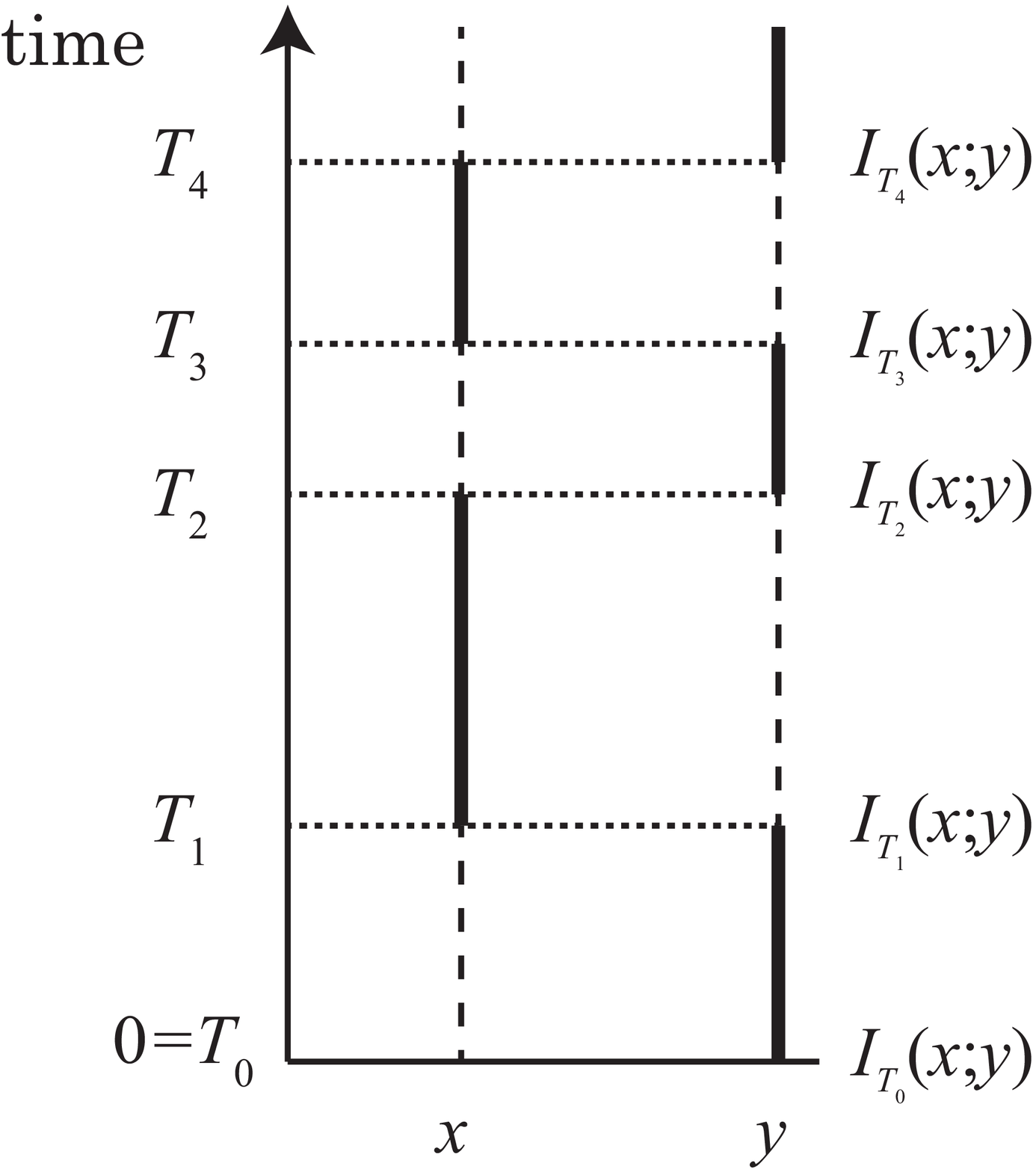}{
Schematic of dynamics of the total system. The bold lines indicate the time intervals when the subsystem can evolve, whereas the dashed lines indicate the time intervals when the subsystem is frozen.
}{f-supple}

We reproduce the FT for non-autonomous Maxwell's demons~\cite{SU2012,SU2012full} from Eq.~\eqref{class2-ss}.
We consider a bipartite system with state $w=(x,y)$.
Intuitively, $x$ is the state of the engine and $y$ is the state of the memory of the demon.
We assume that the transition rates satisfy
\balign{
P(x\to x';t|y)&=0 \ \ (T_{2i}\leq t< T_{2i+1}), \\
&\nt \\
P(y\to y';t|x)&=0 \ \ (T_{2i+1}\leq t< T_{2i+2}),
}
with $0=T_0<T_1<T_2<\cdots <T_{2M}=T$ (See also Fig.~\ref{f-supple}).
In other words, only $y$ can change in time interval $T_{2i}\leq t< T_{2i+1}$, where a measurement is performed by the demon; the measurement outcome is registered in the memory.
Whereas, only $x$ can change in time interval $T_{2i+1}\leq t< T_{2i+2}$, where feedback control is performed; the engine evolves depending on the outcome registered in the memory.

We apply Eq.~\eqref{class2-ss} to this situation and calculate $\Delta I_x$.
While $\Delta I_x$ is equal to zero for time interval $T_{2i}\leq t< T_{2i+1}$,
\eq{
F_x(x,y,t)=\frac{\del}{\del t}I_t(x;y)
}
holds for time interval $T_{2i+1}\leq t< T_{2i+2}$, because the probability distribution $P(x,y,t)$ changes only by transitions in $x$ during this time interval.
Therefore, $\Delta I_x$ for $T_{2i+1}\leq t< T_{2i+2}$ becomes
\balign{
\Delta I_x
=&I_{x,\mr{jump}}+\int_{T_{2i+1}}^{T_{2i+2}}\frac{\del}{\del t}I_t(x;y)dt \nt \\
=&I_{T_{2i+2}}(x;y)-I_{T_{2i+1}}(x;y).
}
We then transform Eq.~\eqref{class2-ss} into
\eq{
\la e^{-\sigma _x+\sum_i I_{T_{2i+2}}(x;y)-I_{T_{2i+1}}(x;y)}\ra =1,
}
which is equivalent to the FT obtained in Refs.~\cite{SU2012,SU2012full}.

\end{document}